\documentclass[conference]{IEEEtran}
\IEEEoverridecommandlockouts
\usepackage{cite}
\usepackage{amsfonts}
\usepackage[cmex10]{amsmath}
\newcommand{\norm}[1]{\left\lVert#1\right\rVert}
\usepackage{amssymb}
\usepackage{algorithmic}
\usepackage{graphicx}
\usepackage{textcomp}
\usepackage{xcolor}

\def\BibTeX{{\rm B\kern-.05em{\sc i\kern-.025em b}\kern-.08em
    T\kern-.1667em\lower.7ex\hbox{E}\kern-.125emX}}

\newcommand{\Exp}{{\mathbb{E}}}
\def\bI{{\mathbf{I}}}
\newcommand{\Ns}{N_{\mathrm{s}}}

\def\bb0{{\mathbb{0}}}

\def\bb{{\mathbf{b}}}

\def\bs{{\mathbf{s}}}

\def\bI{{\mathbf{I}}}

\def\sf0{{\mathsf{0}}}

\def\bsf0{{\bm{\mathsf{0}}}}

\begin{document}

\title{Multidimensional Orthogonal Matching Pursuit-based RIS-aided Joint Localization and Channel Estimation at mmWave\\
}

\author{\IEEEauthorblockN{Murat Bayraktar, Joan Palacios, Nuria Gonz\'alez-Prelcic}\\ 
	\IEEEauthorblockA{%
		\textit{North Carolina State University, USA}\\
		Email:\{\texttt{mbayrak,jbeltra,ngprelcic}\}\texttt{@ncsu.edu}}
\and
\IEEEauthorblockN{Charlie Jianzhong Zhang}
\IEEEauthorblockA{\textit{} \\
\textit{Samsung Research America }\\
Email:\texttt{jianzhong.z@samsung.com}
}}

\maketitle

\begin{abstract}
RIS-aided millimeter wave wireless systems benefit from robustness to blockage and enhanced coverage.
In this paper, we study the ability of RIS to also provide enhanced localization capabilities as a by-product of communication.
We consider sparse reconstruction algorithms to obtain high resolution channel estimates that are mapped to position information.
In RIS-aided mmWave systems, the complexity of sparse recovery becomes a bottleneck, given the large number of elements of the RIS and the large communication arrays. 
We propose to exploit a multidimensional orthogonal matching pursuit strategy for compressive channel
estimation in a RIS-aided millimeter wave system. We show how this algorithm, based on computing the projections on a set of independent dictionaries instead of a single large dictionary, enables high accuracy channel estimation at reduced complexity. We also combine this strategy with a  localization approach which does not rely on the absolute time of arrival of
the LoS path. Localization results in a realistic 3D indoor scenario show that RIS-aided wireless system can also benefit from a significant improvement in localization accuracy.
\end{abstract}


\section{Introduction}

MIMO communication with large arrays and high bandwidths, as used in mmWave bands, enables high data rate communication and also leads to an increased angle and delay resolvability. This, together with the sparse nature of the channel, enhances the sensing capabilities of the communication waveform \cite{ShahmansooriTWC2018}.
Unfortunately, current solutions do not provide the required localization accuracy for some use cases (1cm for indoor and 10 cm for outdoor), even when making unrealistic assumptions or simplifying the evaluation scenarios. 

Reconfigurable intelligent surfaces (RIS) enhance the performance of wireless systems. For example, robustness to blockage is achieved by creating alternative propagation paths which can be controlled  by digitally configuring the coefficients of the passive elements in the RIS. 
From a communication perspective, though, the benefits of RIS  are limited when the line-of-sight (LoS) path is present \cite{BasarAcces2019}. 
If the joint sensing and communication capabilities of a wireless system incorporating RIS are instead considered, the benefits increase. 

RIS can enhance the localization performance with or without the presence of a LoS path between the access point (AP)/base station (BS) and the mobile station (MS), since they create another relatively strong path whose direction can be controlled \cite{HuTSP2018}.
Theoretical lower bounds on the positioning error for RIS-aided localization were obtained in \cite{HuTSP2018,HeVTC2020,ElzanatyTSP2021}, whereas practical positioning algorithms exploiting RIS were proposed in \cite{HeWCNC2020,LinTWC2021,AlbaneseSPAWC2021}. Hierarchical codebooks were designed for RIS phase configurations and training beamformers used to estimate angle of arrival (AoA), angle of departure (AoD) and delay parameters in \cite{HeWCNC2020}. A tensor-based channel estimator was derived in  \cite{LinTWC2021} to extract angle and delay parameters using a two-plane twin RIS structure.  A path parameter estimation approach that exploits the MS position likelihood was 
designed in \cite{AlbaneseSPAWC2021}.
In \cite{HuTSP2018,HeVTC2020,ElzanatyTSP2021,HeWCNC2020,LinTWC2021,AlbaneseSPAWC2021}, 
the pulse shaping / filtering effects at the transmitter and receiver are neglected, leading to an artificial enhancement of  channel sparsity. Furthermore, clock offset between the transmitter and receiver is ignored. Finally, the direct link between the BS and MS is assumed to be blocked, which simplifies the channel estimation task.

In this paper, we propose a  joint channel estimation and localization strategy for  RIS-aided mmWave MIMO systems.
We develop a composite channel model including two potential components (BS-RIS-MS and BS-MS), whose presence depends on the channel realization. We formulate a composite channel estimation problem as a sparse recovery problem with independent dictionaries for the angular and delay domains, and solved with the recently proposed multidimensional orthogonal matching pursuit algorithm (MOMP) \cite{MOMP}. This enables low complexity sparse channel estimation when the large number of antennas and RIS elements prevent the use of a single, large dictionary. To the best of our knowledge, this is the first approach that considers the clock offset in a RIS-aided MIMO system providing position information, and the first one to take into account potential paths from both the RIS and the BS. Simulation results from channels generated by ray tracing show the effectiveness of the approach and the gains in localization accuracy provided by RIS-aided strategies.

\section{System Model}

We consider a frequency-selective mmWave system that incorporates a RIS between the BS and MS to enhance the communication and sensing performance. The BS, RIS and MS are equipped with uniform planar arrays (UPAs), where the number of antennas (passive elements for the RIS) are denoted by $ N_{\mathrm{B}} $, $ N_{\mathrm{R}} $ and $ N_{\mathrm{M}} $, respectively.  Finally, a hybrid analog-digital architecture is employed at the BS and MS, with the number of radio frequency (RF) chains $ N_{\mathrm{RF,B}} $ and $ N_{\mathrm{RF,M}} $.

We focus now on the description of system operation and signal model during initial link establishment. At this stage, the BS and the MS sound the channel aided by the RIS using a set of  $M_{\rm B}$ training transmit configurations  and $M_{\rm M}$ training combiners. A transmit configuration includes a particular choice for the pilot signal, the training precoder and the phase shifts for the RIS. To measure the channel response for all possible combinations of transmit configurations and combiners, the BS  transmits  $M_{\rm B}\times M_{\rm M}$ training frames, each one of them containing $ N_{\mathrm{s}}$ streams of  length-$N$ training sequences. From the received signals corresponding to these training frames, directly from the BS and/or through the RIS, the MS estimates the downlink channel and its own location.
We assume that during training $N_{\mathrm{s}}= N_{\mathrm{RF,B}}$. The training precoder for the $m_{\rm M}$-th training transmit configuration is represented by $ \mathbf{F}_{m_{\rm B}} = \mathbf{F}_{m_{\rm B}}^{\mathrm{RF}} \mathbf{F}_{m_{\rm B}}^{\mathrm{BB}} \in \mathbb{C}^{N_{\mathrm{B}} \times N_{\mathrm{RF,B}}} $, with RF precoder $ \mathbf{F}_{m_{\rm B}}^{\mathrm{RF}} \in \mathbb{C}^{N_{\mathrm{B}} \times N_{\mathrm{RF,B}}} $ and baseband precoder $ \mathbf{F}_{m_{\rm B}}^{\mathrm{BB}} \in \mathbb{C}^{N_{\mathrm{RF,B}} \times N_{\mathrm{RF,B}}} $. Analogously, the $m_{\rm B}$-th training combiner, $m_{\rm B}=1,\ldots,M_{\rm B}$, is denoted by $ \mathbf{W}_{m_{\rm M}} = \mathbf{W}_{m_{\rm M}}^{\mathrm{RF}} \mathbf{W}_{m_{\rm M}}^{\mathrm{BB}} \in \mathbb{C}^{N_{\mathrm{M}} \times N_{\mathrm{RF,M}}} $ with RF combiner $ \mathbf{W}_{m_{\rm M}}^{\mathrm{RF}} \in \mathbb{C}^{N_{\mathrm{M}} \times N_{\mathrm{RF,M}}} $ and baseband combiner $ \mathbf{W}_{m_{\rm M}}^{\mathrm{BB}} \in \mathbb{C}^{N_{\mathrm{RF,M}} \times N_{\mathrm{RF,M}}} $. The training symbol vector at the $n$-th time instance transmitted with the $m_{\rm B}$-th training configuration is denoted as $ \mathbf{s}_{m_{\rm B}}[n] \in \mathbb{C}^{N_{\mathrm{RF,B}}} $, satisfying  ${\Exp}[{\bs}_{m_{\rm B}}[n] {\bs}_{m_{\rm B}}[n]^*] = \frac{1}{\Ns}{\bI}$. To mitigate intersymbol interference, a zero-prefix with length $ D-1 $ is added at the beginning of each frame. $D$ is selected as the delay tap length of the channel.

The training signal generated by the BS is sent through the channel, and can reach the MS directly or via RIS.  This way, the channel model can be written in terms of two components, the BS-MS channel and the BS-RIS-MS channel.  Note that the RIS employs a set of different phase configurations during training, what leads to a different BS-RIS-MS channel matrix for different transmissions.  Mathematically, the overall channel matrix seen when transmitting with the $m_{\rm B}$-th configuration, for the  $ d $-th delay tap, $ d = 0,\dots,D-1 $, can be written as
\begin{equation}
    \mathbf{H}_d^{(m_{\rm B})} = \mathbf{H}_{\mathrm{BM},d} + \mathbf{H}_{\mathrm{BRM},d}^{(m_{\rm B})}, \label{Overall}
\end{equation}
where $ \mathbf{H}_{\mathrm{BM},d} \in \mathbb{C}^{N_{\mathrm{M}} \times N_{\mathrm{B}}} $ is the channel matrix of the BS-MS link and $ \mathbf{H}_{\mathrm{BRM},d}^{(m_{\rm B})} \in \mathbb{C}^{N_{\mathrm{M}} \times N_{\mathrm{B}}} $ is the cascade channel matrix of the BS-RIS-MS link for the RIS matrix corresponding to the the $m_{\rm M}$-th training transmit configuration. 

Leveraging the geometric channel model with $L$ paths, the BS-MS channel can be written as 
\vspace*{-0.2cm}
\begin{multline}
    \mathbf{H}_{\mathrm{BM},d} = \sum_{l=1}^{L} \alpha_{\mathrm{BM},l} \mathbf{a}_{\mathrm{M}}(\boldsymbol{\theta}_{\mathrm{BM},l}) \mathbf{a}_{\mathrm{B}}^H(\boldsymbol{\phi}_{\mathrm{BM},l}) \\
   \times p(dT_s + t_0 - \tau_{\mathrm{BM},l} ), \label{H_BM}
\end{multline}
where each path $ l $ has complex gain $ \alpha_{\mathrm{BM},l} $, AoA $ \boldsymbol{\theta}_{\mathrm{BM},l} $, AoD $ \boldsymbol{\phi}_{\mathrm{BM},l} $, and delay $ \tau_{\mathrm{BM},l} $; the sampling period is denoted as $ T_s $; the time between the beginning of the transmission and the beginning of the reception is $ t_0 $; the time response of the pulse shaping function, which includes the effects of the transmitted signal and the filtering operations at the BS and MS, is represented by $ p(t) $; the array response vectors for the BS and MS are denoted by $ \mathbf{a}_{\mathrm{B}}(\boldsymbol{\phi}) \in \mathbb{C}^{N_{\mathrm{B}}} $ and $ \mathbf{a}_{\mathrm{M}}(\boldsymbol{\theta}) \in \mathbb{C}^{N_{\mathrm{M}}} $ for AoD $ \boldsymbol{\phi} $ and AoA $ \boldsymbol{\theta} $, respectively. We will use vector directions for AoD and AoA instead of polar coordinates, which will be useful in the proceeding operations. Any direction vector $ \boldsymbol{\theta} $ can be expressed as $ \boldsymbol{\theta} = [\theta_x, \theta_y, \theta_z]^T $ with $ \norm{\boldsymbol{\theta}} = 1 $. If a UPA with half the wavelength spacing is placed on the xy-plane, its array response vector can be written as $ \mathbf{a}(\boldsymbol{\theta}) =\mathbf{a}_x(\theta_x) \otimes \mathbf{a}_y(\theta_y) $ with expressions $ [ \mathbf{a}_x(\theta_x) ]_{n_x} =  e^{-j \pi (n_x - 1) \theta_x } $ and $ [\mathbf{a}_y(\theta_y)]_{n_y} =  e^{-j \pi (n_y - 1) \theta_y } $, where $ n_x $ and $ n_y $ are the element indices in x and y directions.

\begin{figure*}[!t]
\normalsize
\begin{equation}
    \mathbf{H}_{\mathrm{BRM},d}^{(m_{\rm B})} = \sum_{q=1}^{Q} \sum_{p=1}^{P} \alpha_{\mathrm{RM},q} \alpha_{\mathrm{BR},p} \mathbf{a}_{\mathrm{M}}(\boldsymbol{\theta}_{\mathrm{RM},q}) \mathbf{a}_{\mathrm{R}}^H(\boldsymbol{\phi}_{\mathrm{RM},q}) \times \boldsymbol{\Omega}^{(m_{\rm B})} \times
    \mathbf{a}_{\mathrm{R}}(\boldsymbol{\theta}_{\mathrm{BR},p}) \mathbf{a}_{\mathrm{B}}^H(\boldsymbol{\phi}_{\mathrm{BR},p}) p(dT_s + t_0 - (\tau_{\mathrm{BR},p}+\tau_{\mathrm{RM},q}) ), \label{Cascade}
\end{equation}
\hrulefill
\vspace*{0pt}
\end{figure*}

The second contribution to the channel in \eqref{Overall}, that represents the BS-RIS-MS link, is described in \eqref{Cascade}, where the diagonal phase reflection matrix at the RIS for the $m_{\rm B}$-th training transmit configuration is represented by $ \boldsymbol{\Omega}^{(m_{\rm B})} = \mathrm{diag} ( \boldsymbol{\omega}^{(m_{\rm B})} ) \in \mathbb{C}^{N_{\mathrm{R}} \times N_{\mathrm{R}}} $, with a phase reflection vector $ \boldsymbol{\omega}^{(m_{\rm B})} = [\omega_1^{(m_{\rm B})}, \cdots, \omega_{N_{\mathrm{R}}}^{(m_{\rm B})}]^T $ which has unit-modulus entries. The number of paths for the BS-RIS and RIS-MS channels are denoted by $ P $ and $ Q $, respectively. It is assumed that the channel BS-RIS is known. The channel parameters (i.e., complex gain, AoA, AoD and delay) for each path are defined in a similar way as for the BS-MS channel. Finally, the array response vector for the RIS is $ \mathbf{a}_{\mathrm{R}}(\cdot) \in \mathbb{C}^{N_{\mathrm{R}}} $. 

With this definition of the channel in mind, the received signal at the MS for the $n$-th time instance and the  $(m_{\rm B},m_{\rm M})$-th training configuration can be written as
\vspace*{-0.2cm}
\begin{multline}
    \mathbf{y}_{m_{\rm B}, m_{\rm M}}[n] = \sqrt{P_t} \mathbf{W}_{m_{\rm M}}^H \sum_{d=0}^{D-1} \mathbf{H}_d^{(m_{\rm B})} \mathbf{F}_{m_{\rm B}} \mathbf{s}_{m_{\rm B}}[n-d]\\
    + \mathbf{W}_{m_{\rm M}}^H \mathbf{v}_{m_{\rm B}, m_{\rm M}}[n], \label{Received}
\end{multline}
for $ n = 1,\dots,N $, where $ P_t $ is the transmit power. $ \mathbf{v}_{m_{\rm B}, m_{\rm M}}[n] \in \mathbb{C}^{N_{\mathrm{RF,M}}} $ is the noise vector for the $ (m_{\rm B}, m_{\rm M}) $-th training configuration and $ n $-th time instance, with independent and identically distributed entries obeying $ \mathcal{NC}(0,\sigma^{2}) $. Note that the noise at the output of the combiner in \eqref{Received} becomes correlated when the combiner is not orthogonal.

\section{Channel Estimation via Multidimensional Orthogonal Matching Pursuit}
In this section we formulate the channel estimation problem using the MOMP algorithm \cite{MOMP}, exploiting independent dictionaries in the angular and delay domain. This algorithm enables a low complexity sparse recovery solution in a scenario where the introduction of the RIS increases the dimensions of the sensing matrix and the final localization application requires of high resolution dictionaries.

\subsection{Formulation of the multidimensional dictionaries}\label{sec:dic}
The MOMP algorithm \cite{MOMP} can recover a  multidimensional sparse signal based on a set of observations and assuming  that it can be represented
by a product of projections on a given set of sparsifying dictionaries. Exploiting the results in \cite{MOMP} is not straightforward, since for the RIS-aided scenario the received signal is a combination of paths coming from two sources, namely the BS-MS channel and the BS-RIS-MS channel. Because of this, we propose exploiting  two sets of dictionaries to model every path.
To further reduce computational complexity, instead of defining separate dictionaries for AoA, AoD and delays for each one of the two channel components, we decide to consider dictionaries for the AoD and for the delay only, so that the AoA information will be embedded into the equivalent complex gain for each path.
Note that the dictionary for AoD can be further decomposed into two dictionaries because a UPA  has an array response vector that can be written as the Kronecker product of two array responses \cite{MOMP}.

Let us define first the dictionaries for the BS-MS channel. In this case, the BS array response is $ \mathbf{a}_{\mathrm{B}}(\boldsymbol{\phi}) = \mathbf{a}_{\mathrm{B},x}(\phi_x) \otimes \mathbf{a}_{\mathrm{B},y}(\phi_y) $, with $ \mathbf{a}_{\mathrm{B},x}(\phi_x) \in \mathbb{C}^{N_{\mathrm{B},x}} $ and $ \mathbf{a}_{\mathrm{B},y}(\phi_y) \in \mathbb{C}^{N_{\mathrm{B},y}} $. The pulse shaping function observed at each delay tap can be stacked in a vector $ \mathbf{a}_{\mathrm{T}}(\tau) \in \mathbb{C}^{D} $ such that $ [\mathbf{a}_{\mathrm{B, T}}(\tau)]_d = p(dT_s - \tau) $ for a given delay $ \tau $.
Therefore, we can define three dictionaries $\boldsymbol{\Psi}_{\mathrm{BM},k}\in\mathbb{C}^{N_{\mathrm{BM},k}^{\rm s}\times N_{\mathrm{BM},k}^{\rm a}}$ that sparsify the BS-MS channel as
\begin{equation}
\begin{aligned}
\boldsymbol{\Psi}_{\mathrm{BM},1} &= \big[\mathbf{a}_{\mathrm{B},x}^*(\overline{\phi}_{\mathrm{BM}, 1, x}),\dots,\mathbf{a}_{\mathrm{B},x}^*(\overline{\phi}_{\mathrm{BM}, N_{\mathrm{BM}, 1}^{\rm a}, x})\big],
\\
\boldsymbol{\Psi}_{\mathrm{BM},2} &= \big[\mathbf{a}_{\mathrm{B},y}^*(\overline{\phi}_{\mathrm{BM}, 1, y}),\dots,\mathbf{a}_{\mathrm{B},y}^*(\overline{\phi}_{\mathrm{BM}, N_{\mathrm{BM}, 1}^{\rm a}, y})\big],
\\
\boldsymbol{\Psi}_{\mathrm{BM},3} &= [\mathbf{a}_{\mathrm{T}}(\overline{\tau}_1), \cdots , \mathbf{a}_{\mathrm{T}}(\overline{\tau}_{N_{\mathrm{BM}, 3}^{\rm a}})],
\end{aligned}
\label{BM_dic}
\end{equation}
where the angles and delays are discretized using some given resolutions.

Analogously, for the BS-RIS-MS channel, the RIS response can be modeled as $ \mathbf{a}_{\mathrm{R}}(\boldsymbol{\phi}) = \mathbf{a}_{\mathrm{R},x}(\phi_x) \otimes \mathbf{a}_{\mathrm{R},y}(\phi_y) $ with $ \mathbf{a}_{\mathrm{R},x}(\phi_x) \in \mathbb{C}^{N_{\mathrm{R},x}} $ and $ \mathbf{a}_{\mathrm{R},y}(\phi_y) \in \mathbb{C}^{N_{\mathrm{R},y}} $.
This leads to the three dictionaries $\boldsymbol{\Psi}_{\mathrm{BRM},k}\in\mathbb{C}^{N_{\mathrm{BRM},k}^{\rm s}\times N_{\mathrm{BRM},k}^{\rm a}}$ that sparsify the BS-RIS-MS channel:
\begin{equation}
\begin{aligned}
\boldsymbol{\Psi}_{\mathrm{BRM},1} &= \big[\mathbf{a}_{\mathrm{R},x}^*(\overline{\phi}_{\mathrm{RM}, 1, x}),\dots,\mathbf{a}_{\mathrm{R},x}^*(\overline{\phi}_{\mathrm{RM}, N_{\mathrm{BRM}, 1}^{\rm a}, x})\big],
\\
\boldsymbol{\Psi}_{\mathrm{BRM},2} &= \big[\mathbf{a}_{\mathrm{R},y}^*(\overline{\phi}_{\mathrm{RM}, 1, y}),\dots,\mathbf{a}_{\mathrm{R},y}^*(\overline{\phi}_{\mathrm{RM}, N_{\mathrm{BRM}, 1}^{\rm a}, y})\big],
\\
\boldsymbol{\Psi}_{\mathrm{BRM},3} &= [\mathbf{a}_{\mathrm{T}}(\overline{\tau}_1), \cdots , \mathbf{a}_{\mathrm{T}}(\overline{\tau}_{N_{\mathrm{BRM}, 3}^{\rm a}})],
\end{aligned}
\label{BRM_dic}
\end{equation}
where the angles and delays fall in a grid of possible values.

\subsection{Compressed Channel Estimation via MOMP}
Our goal is to estimate the channel matrix from a set of observations of the received signal in \eqref{Received}, exploiting the multiple dictionaries defined in Section~\ref{sec:dic}. To this aim, we need 
to write the whitened received training signals in terms of the dictionaries, the sensing matrices (which contain the effect of the training transmit configurations), and the channel coefficients. 

First, to whiten the received signal in \eqref{Received}, we left multiply \eqref{Received} by  $ \mathbf{L}_{m_{\rm M}}^{-1} $, which can be found from the Cholesky decomposition of the noise correlation matrix, i.e., $ \mathbf{L}_{m_{\rm M}} \mathbf{L}_{m_{\rm M}}^H = \mathbf{W}_{m_{\rm M}}^H \mathbf{W}_{m_{\rm M}} $. This way, the whitened received signal $\bar{\mathbf{y}}_{m_{\rm B}, m_{\rm M}}[n]  \in \mathbb{C}^{N_\mathrm{RF,M}}$ is defined as 
\vspace*{-2mm} 
\begin{multline}
    \bar{\mathbf{y}}_{m_{\rm B}, m_{\rm M}}[n] = \sqrt{P_t} \mathbf{L}_m^{-1}\mathbf{W}_{m_{\rm M}}^H \sum_{d=0}^{D-1} \mathbf{H}_{m_{\rm B},}^{(d)} \mathbf{F}_{m_{\rm B}} \mathbf{s}_{m_{\rm B}}[n-d]\\
    + \bar{\mathbf{v}}_{m_{\rm B}, m_{\rm M}}[n], \label{Received_Sparse}
\end{multline}
where  $ \bar{\mathbf{v}}_{m_{\rm B}, m_{\rm M}}[n] = \mathbf{L}_{m_{\rm M}}^{-1}\mathbf{W}_{m_{\rm M}}^H \mathbf{v}_{m_{\rm B}, m_{\rm M}}[n] \in \mathbb{C}^{N_\mathrm{RF,M}} $ is the noise after whitening.

Next, to build the observation matrix, because the channel is frequency-selective, we need to consider all the time instances in a training configuration 
This way, we define the observation for a given training configuration as $ \mathbf{Y}_{m_{\rm B}, m_{\rm M}} = [\bar{\mathbf{y}}_{m_{\rm B}, m_{\rm M}}[1], \cdots, \bar{\mathbf{y}}_{m_{\rm B}, m_{\rm M}}[N]]^{\rm T} \in \mathbb{C}^{N\times N_{\mathrm{RF,M}}} $. The overall noise vector $ \mathbf{V}_{m_{\rm B}, m_{\rm M}} \in \mathbb{C}^{N\times N_{\mathrm{RF,M}}} $ can be defined in a similar form.
Next, we can group all the measurements in a single observation matrix
\begin{equation}
\mathbf{Y} = \left[
\begin{array}{ccc}
\mathbf{Y}_{1, 1} & \cdots & \mathbf{Y}_{1, M_{\rm M}}\\
\vdots & \ddots & \vdots\\
\mathbf{Y}_{M_{\rm B}, 1} & \cdots & \mathbf{Y}_{M_{\rm B}, M_{\rm M}}\\
\end{array}
\right]\in\mathbb{C}^{M_{\rm B} N\times M_{\rm M} N_{\mathrm{RF,M}}}.
\end{equation}
Analogously, we define $\mathbf{V}\in\mathbb{C}^{M_{\rm B} N\times M_{\rm M} N_{\mathrm{RF,M}}}$.

Our next step is to write the observation as the combination of the signal coming from the RIS and the signal coming directly from the BS, using a representation in terms of the corresponding dictionaries defined in \eqref{BRM_dic} and \eqref{BM_dic}. Thus, on one hand, we aim to write the signal corresponding to the BS-MS link as 
\vspace*{-2mm} 
\begin{equation}
\sum_{{\bf i}\in\mathcal{I}_{\rm BM}}\sum_{{\bf j}\in\mathcal{J}_{\rm BM}}[\boldsymbol{\Phi}_{\rm BM}]_{:, {\bf i}}\prod_{k=1}^3[\boldsymbol{\Psi}_{{\rm BM}, k}]_{i_k, j_k}[{\bf C}_{\rm BM}]_{{\bf j}, :},
\label{sparse_model_BM}
\end{equation}
where $\boldsymbol{\Phi}_{\rm BM}$ is the sensing matrix corresponding to the direct link, $\mathbf{C}_\mathrm{BM}\in\mathbb{C}^{\otimes_{d=1}^3N_{\mathrm{BM}, k}^{\rm a}\times M_{\rm M}N_{\rm RF, M}}$ contains the BS-MS channel coefficients, and the multindices variables ${\bf i}\in\mathcal{I}_{\rm BM}=\{(i_1, i_2, i_3)\in\mathbb{N}^3\text{ such that } i_k\leq N_{{\rm BM}, k}^{\rm s}\}$, ${\bf j}\in\mathcal{J}_{\rm BM}=\{(j_1, j_2, j_3)\in\mathbb{N}^3\text{ such that } j_k\leq N_{{\rm BM}, k}^{\rm a}\}$.
On the other hand, our target expression for the signal received via RIS can be written as
\vspace*{-2mm} 
\begin{equation}
\sum_{{\bf i}\in\mathcal{I}_{\rm BRM}}\sum_{{\bf j}\in\mathcal{J}_{\rm BRM}}[\boldsymbol{\Phi}_{\rm BRM}]_{:, {\bf i}}\prod_{k=1}^3[\boldsymbol{\Psi}_{{\rm BRM}, k}]_{i_k, j_k}[{\bf C}_{\rm BRM}]_{{\bf j}, :},
\label{sparse_model_BRM}
\end{equation}
where $\boldsymbol{\Phi}_{\rm BRM}$ is the sensing matrix corresponding to the BS-RIS-MS link, $\mathbf{C}_\mathrm{BRM}\in\mathbb{C}^{\otimes_{k=1}^3N_{\mathrm{BRM}, k}^{\rm a}\times M_{\rm M}N_{\rm RF, M}}$ contains the BS-RIS-MS channel coefficients, and the multindices variables ${\bf i}\in\mathcal{I}_{\rm BRM}=\{(i_1, i_2, i_3)\in\mathbb{N}^3\text{ such that } i_k\leq N_{{\rm BRM}, k}^{\rm s}\}$, ${\bf j}\in\mathcal{J}_{\rm BRM}=\{(j_1, j_2, j_3)\in\mathbb{N}^3\text{ such that } j_k\leq N_{{\rm BRM}, k}^{\rm a}\}$.

With the dictionaries in \eqref{BM_dic} and \eqref{BRM_dic} in mind, we can express $\mathbf{C}_\mathrm{BM}$ and $\mathbf{C}_\mathrm{BRM} $ as the BS-MS and BS-RIS-MS channel coefficients including information about the path complex gains  $\alpha_l$ and the AoA response $[{\bf W}_{m_{\rm M}}]_{:, n_{\rm RF, M}}^{\rm H}{\bf a}_{\rm M}(\theta_l)$ as
\begin{align}\label{eq:exp_C}
[{\bf C}_{\rm BM}]_{{\bf j}, i_{\rm m}}=\left\lbrace\begin{array}{cl}
[\beta_{{\rm BM}, l}]_{i_{\rm m}} & \text{if } \begin{array}{ccc}
\phi_{{\rm BM}, l, x} & = & \overline{\phi}_{\mathrm{BM}, j_1, x} \\
\phi_{{\rm BM}, l, y}^{\rm} & = & \overline{\phi}_{\mathrm{BM}, j_2, y} \\
\tau_{{\rm BM}, l} & = & \overline\tau_{\mathrm{BM}, j_3}
\end{array}\\
0 & \text{otherwise}
\end{array}\right.,\\
[{\bf C}_{\rm BRM}]_{{\bf j}, i_{\rm m}}=\left\lbrace\begin{array}{cl}
[\beta_{{\rm RM}, q}]_{i_{\rm m}} & \text{if } \begin{array}{ccc}
\phi_{{\rm RM}, q, x} & = & \overline{\phi}_{\mathrm{RM}, j_1, x}\\
\phi_{{\rm RM}, q, y} & = & \overline{\phi}_{\mathrm{RM}, j_2, y}\\
\tau_{{\rm RM}, q} & = & \overline\tau_{\mathrm{RM}, j_3}
\end{array}\\
0 & \text{otherwise}
\end{array}\right..
\end{align}
for $\beta_{{\rm BM}, l}, \beta_{{\rm RM}, q}\in\mathbb{C}^{M_{\rm M}N_{\rm RF, M}}$ defined as $[\beta_{{\rm BM}, l}]_{m_{\rm M}N_{\rm RF, M}+n_{\rm RF, M}} = \alpha_{{\rm BM}, l}[{\bf W}_{m_{\rm M}}]_{:, n_{\rm RF, M}}^{\rm H}{\bf a}_{\rm M}(\boldsymbol{\theta}_{{\rm BM}, l})$, $[\beta_{{\rm RM}, q}]_{m_{\rm M}N_{\rm RF, M}+n_{\rm RF, M}} = \alpha_{{\rm RM}, q}[{\bf W}_{m_{\rm M}}]_{:, n_{\rm RF, M}}^{\rm H}{\bf a}_{\rm M}(\boldsymbol{\theta}_{{\rm RM}, q})$.

The final step is finding $\boldsymbol{\Phi}_{\rm BM}\in\mathbb{C}^{M_{\rm B} N\times\otimes_{k=1}^3N_{\mathrm{BM}, k}^{\rm a}}$ and $\boldsymbol{\Phi}_{\rm BRM}\in\mathbb{C}^{M_{\rm B} N\times\otimes_{k=1}^3N_{\mathrm{BRM}, k}^{\rm a}}$ satisfying the models in \eqref{sparse_model_BM} and \eqref{sparse_model_BRM} to reconstruct the received signal.
We simplify this step by assuming that the channel between the BS and the RIS has a predominant line of sight component, so we can reduce the number of paths to $P=1$ without much loss in performance.
This leads to the expression
\begin{align}\label{eq:exp_A_BM}
[{\bf \Phi}_{\rm BM}]_{m_{\rm B}N+n, {\bf i}} = [{\bf F}_{m_{\rm B}}{\bf s}[n-i_3]]_{i_1N_{\rm T}^{\rm y}+i_2},\\
[{\bf \Phi}_{\rm BRM}]_{m_{\rm B}N+n, {\bf i}} = [\bar{\bf F}_{m_{\rm B}}{\bf s}[n-i_3]]_{i_1N_{\rm T}^{\rm y}+i_2}.
\end{align}
for $\bar{\bf F}_{m_{\rm B}} = \alpha_{\mathrm{BR},1}\boldsymbol{\Omega}^{(m_{\rm B})}
    \mathbf{a}_{\mathrm{R}}(\boldsymbol{\theta}_{\mathrm{BR},1}) \mathbf{a}_{\mathrm{B}}^H(\boldsymbol{\phi}_{\mathrm{BR},1}){\bf F}_{m_{\rm B}}$.

\begin{figure*}[!t]
\normalsize
\begin{equation}
\min_{{\bf C}_{\rm BM}, {\bf C}_{\rm BRM}}\|{\bf Y}-\left(\sum_{{\bf i}\in\mathcal{I}_{\rm BM}}\sum_{{\bf j}\in\mathcal{J}_{\rm BM}}[\boldsymbol{\Phi}_{\rm BM}]_{:, {\bf i}}\prod_{k=1}^3[\boldsymbol{\Psi}_{{\rm BM}, k}]_{i_k, j_k}[{\bf C}_{\rm BM}]_{{\bf j}, :}+\sum_{{\bf i}\in\mathcal{I}_{\rm BRM}}\sum_{{\bf j}\in\mathcal{J}_{\rm BRM}}[\boldsymbol{\Phi}_{\rm BRM}]_{:, {\bf i}}\prod_{k=1}^3[\boldsymbol{\Psi}_{{\rm BRM}, k}]_{i_k, j_k}[{\bf C}_{\rm BRM}]_{{\bf j}, :}\right)\|
\label{eq:MOMP_tweak}
\end{equation}
\hrulefill
\vspace*{0pt}
\end{figure*}

Now, with all these definitions, we can write the channel estimation problem as \eqref{eq:MOMP_tweak}.
We modify the MOMP algorithm in \cite{MOMP} to operate with this alternative formulation by repeating he projection step at each step for both sources independently, and then selecting the best matching projection one.
After getting the MOMP output we extract the paths information from the coefficients support $\hat{\bf j}\in\mathcal{C}_{\rm BM}$ and $\hat{\bf j}\in\mathcal{C}_{\rm BRM}$ like $\hat{\phi}_{x}=\overline{\phi}_{j_1,x}, \hat{\phi}_{x}=\overline{\phi}_{j_2,y}$ and $\hat{\tau}-t_0=\overline{\tau}_{j_3}$ for the corresponding discretized domains.
$\hat{\phi}_{z}$ can be resolved from $\hat{\phi}_{z} = \sqrt{(\hat{\phi}_{x})^2+(\hat{\phi}_{y})^2}$.

\section{Localization}
We consider two different geometric approaches for localization depending on whether there are one or two LoS paths. The first case applies to the classical BS-MS setting without a RIS and to the case with a RIS where only RIS-MS link is LoS. The second case applies only to the RIS setting. Our derivations exploit useful properties that are found in the indoor propagation environment to solve for the unknown propagation offset and solve the localization problem. 

\subsection{Localization with one LoS path}
We begin by describing the localization algorithm by using the BS-MS link. Let $ \mathbf{b} \in \mathbb{R}^3 $ and $ \mathbf{m} \in \mathbb{R}^3 $ be the locations of the BS and MS, and $c$ the speed of light. Then, the MS location is 
\begin{equation}
    \mathbf{m} = \mathbf{b} + c\tau_{\mathrm{BM},1}\boldsymbol{\phi}_{\mathrm{BM},1}, \label{MS_pos}. 
\end{equation}
Unfortunately, the channel estimation algorithm provides only relative delays or time difference of arrival (TDoA) for the paths (i.e. $ \tau_{\mathrm{BM},l} - t_0 $) due to the unknown clock offset $t_0$.  We will use the relative delay and the implication assumption of an indoor localization scenario, where the reflection surfaces are either horizontal or vertical to solve this problem.

Let us classify the non-line-of-sight (NLoS) paths as first-order wall or floor/ceiling reflections. If it is a floor/ceiling reflection, the azimuth angle of that path should be equal to the azimuth angle of the LoS path. This requires that 
\begin{equation}
    \frac{ \phi_{\mathrm{BM},l,x} \phi_{\mathrm{BM},1,x} + \phi_{\mathrm{BM},l,y} \phi_{\mathrm{BM},1,y}}{ \sqrt{\phi_{\mathrm{BM},l,x}^2+\phi_{\mathrm{BM},l,y}^2} \sqrt{\phi_{\mathrm{BM},1,x}^2+\phi_{\mathrm{BM},1,y}^2} } = 1.
\end{equation}
\noindent
For such paths, a floor/ceiling path travels the same horizontal distance as the LoS path, satisfying 
\begin{multline}
    \sqrt{\phi_{\mathrm{BM},l,x}^2+\phi_{\mathrm{BM},l,y}^2} ((\tau_{\mathrm{BM},l} - t_0) + t_0) \\ = \sqrt{\phi_{\mathrm{BM},1,x}^2+\phi_{\mathrm{BM},1,y}^2} ((\tau_{\mathrm{BM},1} - t_0) + t_0). \label{Floor}
\end{multline}
\noindent
Similarly, if a NLoS path is reflected by a wall, the vertical distance traveled by the LoS and NLoS path is the same, satisfying
\begin{equation}
    \phi_{\mathrm{BM},l,z} ((\tau_{\mathrm{BM},l} - t_0) + t_0) = \phi_{\mathrm{BM},1,z} ((\tau_{\mathrm{BM},1} - t_0) + t_0). \label{Wall}
\end{equation}
\noindent
Paths that do not satisfy the ceiling or wall conditions are discarded from the path pool. After classifying the NLoS paths, the estimate $\hat{t}_0$ is found using the set of equations given by \eqref{Floor} and \eqref{Wall}. Then the true $\tau_{\mathrm{BM},1}$ is computed and the MS position is estimated using \eqref{MS_pos}.

While we described the localization for the BS-MS link, a similar approach also works for the case where the BS-MS link is obstructed and there exists a LoS path for the RIS-MS link. Let the position of the RIS be denoted by $ \mathbf{r} \in \mathbb{R}^3 $. Then, we can obtain an estimate of the MS position with the aid of RIS, if we replace $ \mathbf{m} $, $ \tau_{\mathrm{BM},l} $ and $ \boldsymbol{\phi}_{\mathrm{BM},l} $ with $ \mathbf{r} $, $ \tau_{\mathrm{RM,q}} $ and $ \boldsymbol{\phi}_{\mathrm{RM,q}} $, respectively, in the equations above.

\subsection{Localization with two LoS paths}
Now we assume that LoS paths exist for both the BS-MS and RIS-MS links. In this case, it is possible to find the location of the MS by using just the two LoS paths; the other NLoS paths are not required. The key idea is that the user position computed from \eqref{MS_pos} or the RIS-MS equivalent is the same:
\begin{equation}
\begin{aligned}
    \mathbf{m} &= \mathbf{b} + c((\tau_{\mathrm{BM},1} - t_0) + t_0)\boldsymbol{\phi}_{\mathrm{BM},1} \\
    &= \mathbf{r} + c((\tau_{\mathrm{RM},1} - t_0) + t_0)\boldsymbol{\phi}_{\mathrm{RM},1},\label{MS_pos2}
\end{aligned}    
\end{equation}
\noindent
Using the estimated TDoAs, then \eqref{MS_pos2} is a simple linear equation in the unknown $t_0$ which can be solved accordingly. Then the estimate of $ t_0 $ can be substituted in either of the equations to find the MS location. It is meaningful to use the LoS path that has a higher gain since it could possibly provide more accurate estimates.

\section{Numerical Results}
We consider an indoor factory environment with dimensions $ 60\times120\times10 \mathrm{m^3} $ with the origin taken as the middle point on the bottom of the North wall. There are several wooden and metal boxes on the floor with different sizes. The center frequency is set to $ 60\mathrm{GHz} $ while the bandwidth is $ 100\mathrm{MHz} $. The transmit power is set to $ P_t = 20\mathrm{dBm} $, whereas the noise variance is $ \sigma^2 = -94\mathrm{dBm} $, which is the thermal noise at $15^\circ\mathrm{C}$ with the given bandwidth. The pulse shaping function is selected as $ p(t) = \mathrm{sinc}(t) $. The delay tap length and the number of training symbols are set to $ D = 32 $ and $ N = 64 $, respectively. Training symbols are selected as the rows of the $ 64 $ element Hadamard matrix. The dictionaries have a high resolution such that the ratio of the number of columns to rows of each dictionary is set to $ 128 $. The BS and MS are equipped with $ 8\times8 $ UPA with $8$ RF chains, and $ 4\times4 $ UPA with $4$ RF chains, respectively. The distance unit, which is $ \mathrm{m} $, is omitted for the rest of the parameters. The BS is horizontally placed on the ceiling with center $ \mathbf{b} = [10,-10,9.5]^T $, whereas the RIS is vertically mounted on the wall with center $ \mathbf{r} = [0,0,5.5]^T $. We randomly deploy 100 MSs with arrays located horizontally at coordinates $ m_x \in [-10,0] $, $ m_y \in [-15,-5] $ and $ m_z = 1.5 $. The paths for the BS-MS, BS-RIS and RIS-MS channels are generated via a ray-tracing software.

%
\begin{figure}[htbp]
\centerline{\includegraphics[width=0.8\columnwidth]{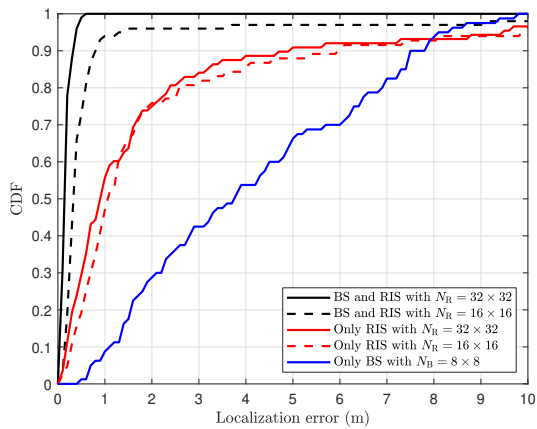}}
\caption{Localization error with only BS-MS, only RIS-MS, or both channels.}
\label{Fig2}
\end{figure}

We study  the empirical CDF of the localization error, shown in~Fig. \ref{Fig2}. The precoders and combiners are randomly generated for the case with only the BS-MS link. The columns of the training precoders are set to the array response corresponding to the AoD of the LoS path between BS-RIS (i.e., $ \mathbf{a}_{\mathrm{B}}^H(\boldsymbol{\phi}_{\mathrm{BR},1}) $), and the reflection matrices are randomly generated for the only RIS-MS link.
The number of training frames is set to the half the number of antenna/passive elements for all cases. For the case with both the BS-MS and RIS-MS links, we set the half of the columns of the precoder to $ \mathbf{a}_{\mathrm{B}}^H(\boldsymbol{\phi}_{\mathrm{BR},1}) $, while the other columns, combiners and phase reflection matrices are randomly generated.  RIS-aided localization clearly outperforms localization without RIS. 
The scenarios where LoS paths of both the BS-MS and RIS-MS channels are available provides the lowest localization error. Introducing a $32\times32$ RIS guarantees an error smaller than 20 cm for 80\% of the user positions in the data set, while the localization error without RIS is only smaller than 6.5 m for the same percentage of users. These results clearly show the significant  improvement on position accuracy that RIS can provide.

\section{Conclusion}
We developed a low complexity compressive channel estimation strategy for a RIS-aided mmWave system, leveraging the 
MOMP algorithm. We integrated this approach with a new localization strategy that can operate with or without the RIS.
We generated a set of realistic indoor channels using ray tracing, and we showed the significant improvement in localization accuracy that the RIS can provide, even without perfect synchronization assumptions. 

\bibliographystyle{IEEEtran} 
\bibliography{References.bib} 

\end{document}